# MOON'S RADIATION ENVIRONMENT AND EXPECTED PERFORMANCE OF SOLAR CELLS DURING FUTURE LUNAR MISSIONS

T.E Girish and S Aranya
Department of Physics,
University College,
Thiruvananthapuram 695034, INDIA
Email: tegirish5@yahoo.co.in

**Abstract**

Several lunar missions are planned ahead and there is an increasing demand for efficient photovoltaic power generation in the moon. The knowledge of solar cell operation in the lunar surface obtained during early seventies need to be updated considering current views on solar variability and emerging space solar cell technologies. In this paper some aspects of the solar cell performance expected under variable lunar radiation environment during future space missions to moon are addressed. We have calculated relative power expected from different types of solar cells under extreme solar proton irradiation conditions and high lunar daytime temperature. It is also estimated that 2-3 % of annual solar cell degradation is most probable during the future lunar missions. We have also discussed photovoltaic power generation in long term lunar bases emphasizing technological needs such as sunlight concentration, solar cell cooling and magnetic shielding of radiation for improving the efficiency of solar cells in the lunar environment.

## 1. Introduction

Thirty years have passed since the early solar cell experiments in moon as a part of Apollo [1] and Lunokhod missions [2]. The space solar cell technology has significantly evolved during this period.[3].There is now a renewed interest in lunar exploration with intentions to tap the extensive energy and mineral resources present there [4] .Understanding and prediction of the lunar radiation environment is essential for efficient photovoltaic power generation in moon. Its important characteristics and the observed solar cycle variability [ 5-7] is summarized in Table.1

In this paper we have addressed some aspects of solar cell performance expected during future lunar missions [8] under severe conditions of the lunar radiation environment. The study is carried out using our current knowledge solar variability and radiation resistance of solar cells obtained from previous laboratory/ space born experiments. We have calculated relative power expected from different solar cells under extreme proton irradiation conditions in moon. The annual solar cell degradation in the lunar radiation environment is also estimated for use in future lunar missions considering different factors contributing to the same. Finally we have also discussed methods of ensuring efficient photovoltaic power generation in long term lunar bases by adopting techniques such as concentrated photovoltaic systems, solar cell cooling and magnetic shielding of solar proton radiation .

## 2. Solar variability and severe solar proton events.

Energy (E) and fluence (F) of solar proton events ( SPE's )are available from direct (using satellites) and indirect ( eg. ionospheric effects ) observations for the past five sunspot cycles [9-11]. The annual occurrence of SPE's (N) with E> 10MeV along with yearly mean sunspot number ( R) for the years 1955-2006 is plotted in Figure.1. We could obtain a statistically significant correlation between R and N ( r = 0.62) during this period. This result implies that from the knowledge of the probable time of occurrence of future sunspot maxima [12] we can predict period of enhancement in the occurrence number of SPE's.The fluence of solar proton events however show irregular changes and cannot be easily predicted.

The period of occurrence of large fluence solar proton events (> 30MeV, fluence >$2X10^9$ cm$^{-2}$) has been inferred for the past 400 years ( 1561-1950 AD) using nitrate deposition in polar ice cores [ 13].During the long interval from 1561 to 2007 AD the most severe solar proton event is inferred during September 1859 [ 14-15] in association with the historic Carrington solar flare event and occurrence of a super intense geomagnetic storm. The characteristic of this most severe SPE are the following :

E> 30MeV , F = 18.8 X $10^9$ cm$^{-2}$ .This can be considered as an extreme limit of the future solar proton fluence which may occur during future lunar mission periods.

## 3. Expected performance of different solar cells in moon under extreme solar proton irradiation and lunar temperature

Extensive data is available on the radiation resistance of different solar cells subjected to high energy proton radiations under laboratory and space-born conditions. In Table.2 we have shown our calculations of net reduction in efficiency and relative power output per unit area of different solar cells under the extreme limit of SPE's (defined in section 2.) and maximum lunar day time temperature.. The calculations are carried out for Si ,GaAs, multi junction Group III- V and CIGS solar cells [ 16-21]. For convenience the power out put from Si solar cells at moon is taken as unity. Also for different materials, solar cell efficiency during initial conditions are assumed to be identical. Maximum relative power output is found for multi junction Group III – V solar cells and , minimum power for Si solar cells.

## 4. Predicting long term solar cell degradation in the lunar radiation background.

In the previous section we have discussed the possible effect of a single and severe SPE event on the out put of solar cells made of different materials . We are also interested in knowing the long term degradation of solar cells in the lunar radiation background. In Table.3 we have shown , average solar output decrease observed per year for four different space missions, including the Apollo 14/15 solar cell experiments in moon

[ 3, 20,21-23]. The number of large fluence solar proton events (F > $2 \times 10^9$ cm$^{-2}$) observed during these mission periods are also shown in this table. The observed solar cell degradation of 2-3% per year is expected to be applicable for future long term space missions to moon.

## 5. Efficient Photovoltaic Power generation for long term lunar bases

Photovoltaic systems can partially meet the power needs of a long term lunar base [24]. Manufacturing solar cells in moon, making use of lunar materials is a viable option for this purpose[ 25] Silicon is abundant in moon and Si solar cell technology [26] is easy to implement. The low radiation resistance and large temperature dependent degradation of Si solar cells is a matter of concern. However, it may be possible to overcome these defects of Si solar cells and ensure efficient photovoltaic power generation in moon by adopting one or more of the following techniques:
( i ) PV concentrated systems (ii) solar cell cooling and (iii) magnetic shielding of lunar radiation.

Adopting active or passive cooling techniques [27-28] the high solar cell temperature at lunar noon time (about $100^0$C) can be brought down to ambient conditions in earth (25ºC) which can improve the efficiency of Si solar cells at moon by 40% as seen from Table.2. Alternatively one can use concentrator photovoltaic systems for power generation during the lunar day.Under concentrated sunlight (X 100) Si solar cells when subjected to severe proton irradiation(37MeV, $10^{10}$ particles/cm$^2$), the solar cell efficiency decreased only by 10% at a cell temperature of $80^0$ C [16]. However, extra cooling arrangements are preferable for photovoltaic power generation during lunar noon time with concentrated systems.

Cooling of Si solar cells below 90K during the lunar day using cryogenic techniques has several advantages. LILT Solar cell operation with enhanced efficiency may become feasible [29] Since helium isotopes are abundant in the lunar regolith and nitrogen is present in lunar soil [30] production of cryofluids such as liquid nitrogen and liquid helium may be feasible in moon. Magnetic shielding of harmful lunar radiations becomes essential for manned lunar bases. . This can protect not only life but also solar cells i from severe solar proton irradiation. For magnetic shielding of protons [31] within a spherical region of radius $r_p$ in moon is governed by the equation

$$r_p \text{ (metres)} = \text{SQRT}(kM/p) \qquad (1)$$

$$\text{Here } k = (\mu q/4\pi)$$

Where q is proton charge, $\mu$ is permeability of free space, M is the moment of the electromagnet used for radiation shielding and p is the momentum of the solar proton. It is estimated that for protons of 200MeV energy we require a magnetic moment of $10^{10}$ $Am^2$ for radiation shielding. Superconducting magnets with Tc lying in the liquid nitrogen temperature range with critical current density of $10^6$ $A/cm^2$ can be used for this purpose. Si solar cells placed in a circular tank of liquid nitrogen over the roof of the lunar base can be protected from severe solar proton radiations by employing such high Tc superconducting wire magnets.

## 6. Discussion

The prediction of fluence of solar proton events is a difficult problem [32]. The fluence of SPE's observed in connection with the Carrington solar flare event of September 1859 can be considered as an upper limit ($1.8 \times 10^{10}$ particles/$cm^2$) of SPE's probably encountered during future lunar missions. The performance of Si, GaAs, multijunction Gr. III-V and CIGS solar cells under this extreme proton irradiation (>30Mev) is summarized in Table. 2 ,where temperature related degradation in moon is also taken into account. We have also estimated the relative power output expected from this solar cells in moon. This is found to be maximum for multijunction Gr. III-V solar cells with a value 2.4 times that of Si solare cells. . The GaAs solar panels used in Lunokhod Lunar rovers during early seventies had an average efficiency of 11% in moon and their power output is found to be two times that Si solar cells tested under identical conditions [3]. It is interesting to note that our estimate of the relative power output of GaAs solar cells given in Table 2 matches with the lunokhod space mission results.

Long term solar cell degradation in near earth space environments or moon is controlled by by different factors such as SPE's, galactic cosmic rays UV/X ray radiations from the Sun and thermal recycling ( due to dayside and nightside temperature differences) effects. The contribution from SPE's is the dominant factor affecting solar cell performance in moon. The fluence of galactic cosmic ray particles is at least two orders of magnitude less than that of solar energetic protons [33]. Laboratory studies of solar UV radiation exposures of Si solar cells suggested 1.5-2 % degradation of solar cell output per year [34]. One will be able to minimize thermal recycling effects of solar cells [2] in moon by adopting solar cell cooling techniques.

Large fluence solar proton events like that occurred during 1972 August, 1989 March, etc can cause a solar cell degradation equivalent to that of 0.5 to 1 year exposure to space radiation environment [22]. .Thus the occurrence number of major SPE'S with $F > 2 \times 10^9$ cm$^{-2}$ during a space mission period becomes important to calculate the EOL efficiency of solar cells. The maximum frequency of occurrence of such large fluence SPE's per sunspot cycle can be up to six to eight [15]. For a lunar mission lasting 5-6 years we can expect up to a maximum of four major SPE events similar to that occurred during the years 2000-2005 [16].. The Gr III-V multijuncion solar cell arrays ( AlGa As/Ga As) in the MIR space station [3] showed only a degradation of 30 % at the end of fifteen years of operation ( 1986-2000). Thus our estimate of 2-3 % solar cell degradation per annum to seem be most probable for future lunar missions.

Since the energy, fluence and composition of galactic cosmic rays(GCR) observed is complex and time varying it is not easy to simulate GCR radiations in the laboratory. During daytime GCR effects in interplanetary space are combined with contributions from solar wind ( including SPE's) ,solar UV radiation etc.Moon offers a natural laboratory during its prolonged night periods ( 14.5 days) for studying GCR effects on technological systems such as solar cells when the contributions from other radiation sources are nearly absent. We can plan as a part of future lunar missions,solar cell degradation experiments during lunar night. Such solar cells can be illuminated by steady artificial sources such as a laser powered by a battery or reflected albedo radiation from the earth [29].

Cost , technology and lunar abundance are the deciding factors for the choice of solar cell materials for photovoltaic power generation in the long term lunar bases. Sunlight concentration and solar cell cooling arrangements may be necessary for efficient photovoltaic power generation if we employ Si and CIGS solar cells. Solar cell cooling making use of liquid nitrogen has certain additional advantages. For manned lunar bases radiation shielding of solar cells as well as living organisms in the lunar base can be done simultaneously by using high Tc , YBCO superconducting wire electromagnets [ 35] kept at liquid nitrogen temperatures ( 77 K ). But the effect of strong magnetic fields on the photovoltaic power generation has to be considered in this context. Earlier reports on the solar cell operation in the presence of external magnetic fields will be helpful in this context [36-37].

**6.Conclusions**

(i). We have calculated the relative power output expected from different types of solar cells in moon when operated during lunar day under extreme solar proton event conditions and high lunar temperature.

(ii) Considering different factors contributing to the long term solar cell degradation in the lunar environment ,2-3 % decrease per year seem to be most probable during the future lunar missions.

(iii) For efficient photovoltaic power generation in moon using lunar manufactured solar cells , techniques such as sunlight concentration with solar cell cooling and magnetic shielding of radiation can be considered.

**Aknowledgements**

The authors wish to thank Prof.Jacob Koshy and Mr.Abilash Kumar for helpful comments. One of the authors ( S.Aranya) is grateful to the University of Kerala for the award of a Junior Research Fellowship.

**Table 1: Lunar radiation Environment and solar cycle variability**

| Type of radiation | relative contribution to total at moon | energy range | solar cycle change |
|---|---|---|---|
| Protons | 80% | 1-100 MeV | (SPE occurrence) X 15 |
| Neutrons | 18% | MeV- GeV | X 3 |
| Other particles and nuclei | 2% | MeV- GeV | X 3 |
| EUV/Xray flux from Sun | 0.1% | KeV-MeV | X 10 |

**Table 2: Solar cell efficiency at moon under extreme SPE's and high lunar temparature**

| Type of the solar cell | Decrease due to proton radiation E > 30 MeV Fluence $10^{10}$ /cm$^2$ | Decrease due to high temperature ( up to 100°C) | Net decrease estimated | Relative power estimated at moon |
|---|---|---|---|---|
| Si | 20 -25% | 40-45% | 65% | 1 |
| GaAs | 5-10% | 20-25% | 30% | 2 |
| Multijunction/ Tandem Group III –V | 5-10% | 10% | 20% | 2.3 |
| CIGS | 0% | 50% | 50% | 1.4 |

**Table. 3 : Average solar cell degradation in per year calculated for different space missions.**

| Period | Space mission | No. of major SPE's occurred | Average solar cell output degradation per year |
|---|---|---|---|
| 1971-76 | Apollo-14/15 (Lunar) | 1 | 2% (Si) |
| 1987-97 | ETS V (Earth-Orbiter) | 1 | 2.5% (Si), 3% (GaAs) |
| 1986-2000 | MIR station | 2 | 2% (AlGa As/Ga As) |
| 2001-05 | CLUSTER (Earth – Orbiter) | 3 | 3% |

**Figure Captions**

Fig. 1. Variations in the yearly occurrences of solar proton events (SPE's) with energy greater than 30 MeV and annual mean sunpot number for the years 1955-2006.

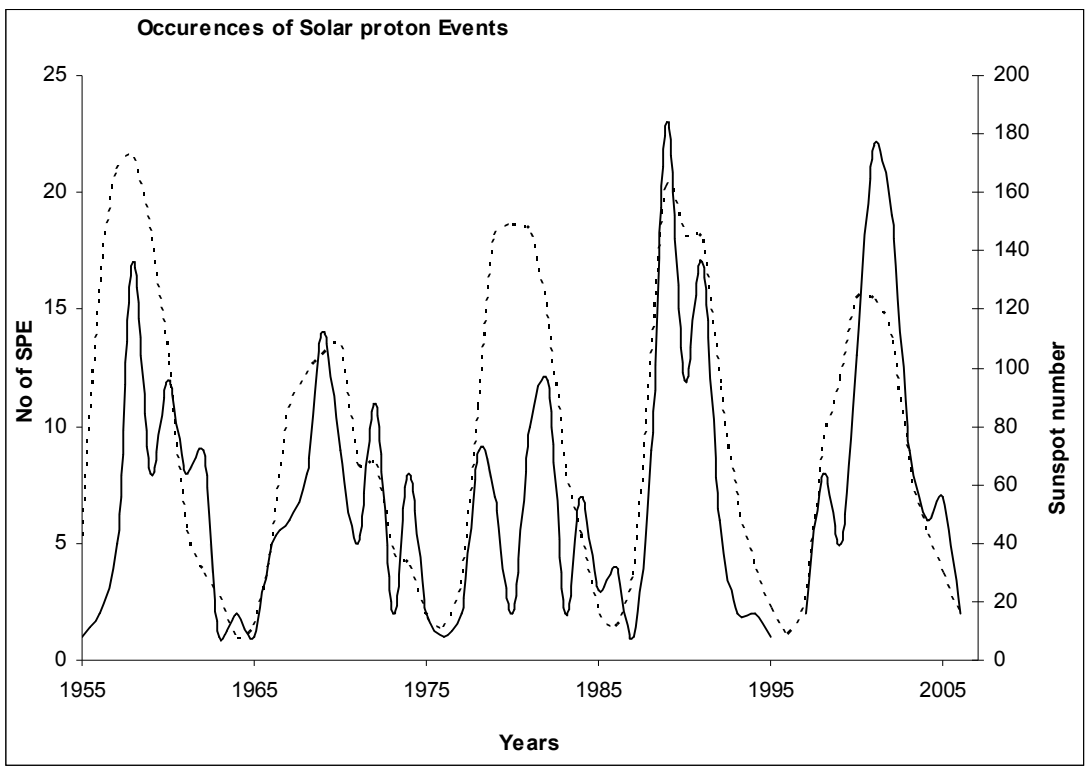

**Fig.1**

```
 1
 2
 3
 4
 5
 6
 7
 8
 9
10
11
12
13
14
15
16
17
18
19
20
21
22
23
24
25
26
27
28
29
30
31
32
33
34
35
36
37
38
39
40
41
42
43
44
45
46
47
48
49
50
51
52
53
54
55
56
57
58
59
60
61
62
63
64
65
```